\documentclass[aps,pre,twocolumn,groupedaddress,showpacs]{revtex4-1}
\usepackage{amsmath,amssymb,subfigure,fancybox,txfonts,bm,color,wrapfig}
\usepackage{siunitx,numprint}
\usepackage[dvipdfmx]{graphicx}
\usepackage{bm}      % bold math
\usepackage{braket}

\begin{document}

\newcommand{\vc}{\mathbf}
\newcommand{\gvc}[1]{\mbox{\boldmath $#1$}}
\newcommand{\fracd}[2]{\frac{\displaystyle #1}{\displaystyle #2}}
\newcommand{\ave}[1]{\left< #1 \right>}
\newcommand{\red}[1]{\textcolor[named]{Red}{#1}}
\newcommand{\blue}[1]{\textcolor[named]{Blue}{#1}}
\newcommand{\green}[1]{\textcolor[rgb]{0,0.6,0}{#1}}
\newcommand{\del}[3] {\frac{\partial^{#3} #1}{\partial #2^{#3}}}
\newcommand{\dev}[3]{\frac{\text{d}^{#3} #1}{\text{d}#2^{#3}}}
\newcommand{\pdev}[3]{{\text{d}^{#3} #1}/{\text{d}#2^{#3}}}
\newcommand{\pdel}[3]{{\partial^{#3} #1}/{\partial #2^{#3}}}
\newcommand{\intd}[1]{\text{d} {#1}}
\newcommand{\emf}[1]{{\gtfamily \bfseries #1}}
\newcommand{\subti}[1]{\begin{itemize} \item \emf{ #1} \end{itemize}}

\newcommand{\Real}{\operatorname{Re}}
\newcommand{\Imag}{\operatorname{Im}}

\newcommand{\am}{{\bm a}}
\newcommand{\bb}{{\bm b}}
\newcommand{\ff}{{\bm f}}
\newcommand{\pp}{{\bm p}}
\newcommand{\rr}{{\bm r}}
\newcommand{\sm}{{\bm s}}
\newcommand{\tm}{{\bm t}}
\newcommand{\uu}{{\bm u}}
\newcommand{\ww}{{\bm w}}
\newcommand{\xx}{{\bm x}}
\newcommand{\yy}{{\bm y}}
\newcommand{\zz}{{\bm z}}
\newcommand{\Model}{{\mathbb M}}
\newcommand{\RR}{{\mathbb R}}
\newcommand{\NN}{{\mathbb N}}
\newcommand{\oomega}{\mbox{\boldmath $\omega$}}
\newcommand{\WW}{{\bm W}}
\newcommand{\EE}{\mbox{\boldmath $E$}}
\newcommand{\FF}{\mbox{\boldmath $F$}}
\newcommand{\KK}{\mbox{\boldmath $K$}}
\newcommand{\GG}{\mbox{\boldmath $G$}}
\newcommand{\tr}{\mathrm{T}}
\newcommand{\CC}{\mbox{$\hat{C}$}}
\newcommand{\II}{\mbox{$\hat{I}$}}
\newcommand{\HH}{\mbox{$\hat{H}$}}
\newcommand{\MM}{\mbox{$\hat{M}$}}
\newcommand{\Am}{\mbox{$\hat{A}$}}
\newcommand{\PP}{\mbox{$\hat{P}$}}
\newcommand{\QQ}{\mbox{$\hat{Q}$}}
%\usepackage{amsmath}	% required for `\align' (yatex added)
%
%\begin{document}

\title{Gigahertz-rate random speckle projection for high-speed single-pixel image classification \\
}

\author{Jinsei Hanawa$^1$}
\author{Tomoaki Niiyama$^2$}
\author{Yutaka Endo$^2$}
\author{Satoshi Sunada$^{2,3}$}
\email{sunada@se.kanazawa-u.ac.jp}

\affiliation{
$^1$Graduate School of Natural Science and Technology, Kanazawa University, 
Kakuma-machi, Kanazawa, Ishikawa 920-1192, Japan \\
$^2$Faculty of Mechanical Engineering, Institute of Science and
Engineering, Kanazawa University\\
Kakuma-machi, Kanazawa, Ishikawa 920-1192, Japan \\
$^3$Japan Science and Technology Agency (JST), PRESTO, 4-1-8 Honcho,
 Kawaguchi, Saitama 332-0012, Japan\\
}

\begin{abstract}
Imaging techniques based on single-pixel detection, such as ghost imaging, can reconstruct 
or recognize a target scene from multiple measurements using a sequence 
of random mask patterns. 
However, the processing speed is limited by the low rate of the pattern generation.
In this study, we propose an ultrafast method for random speckle pattern generation, which has the potential to overcome the limited processing speed.
The proposed approach is based on multimode fiber speckles induced by 
fast optical phase modulation.
We experimentally demonstrate 
dynamic speckle projection with phase modulation at 
10 GHz rates, which is five to six orders of magnitude higher than 
conventional modulation approaches using spatial light modulators.
Moreover, we combine the proposed generation approach with 
a wavelength-division multiplexing technique and 
apply it for image classification.
As a proof-of-concept demonstration, we show that
28$\times$28-pixel images of digits acquired at GHz rates 
can be accurately classified using a simple neural network.
The proposed approach opens a novel pathway for an all-optical image processor.
\end{abstract}

\maketitle
%%%%%%%%%%%%%%%%%%%%%%%%%%  body  %%%%%%%%%%%%%%%%%%%%%%%%%%
\section{Introduction \label{sec1}}
Random pattern generation has garnered considerable interest because it has a wide range of potential applications, including imaging technologies based on single-pixel detection \cite{Gibson:20,Higham:2018aa,4472247,doi:10.1126/science.1234454}, such as ghost imaging \cite{PhysRevA.79.053840,Ye:20,Chen:09}, as well as remote sensing \cite{Erkmen:12} and photonic computing \cite{Sunada:20,7472872,9527089}.
Spatial light modulators (SLMs) have been commonly used to generate optical random patterns.
While SLMs with a large number of pixels enable a wide variety of illumination pattern controls, the low refresh rate, which is typically on the order of $\sim$10 kHz, ultimately limits the image processing speed. 
In response, significant effort has been devoted to increasing the refresh rate of the pattern illumination for high-speed ghost imaging and single-pixel image processing \cite{Hahamovich:2021aa,Xu:18,wang2018ultrafast,Shi:21,Fukui:21_Lightwave,doi:10.1063/1.5001750,Wang:2017ab}.

Owing to the high speed wavelength-to-time mapping and modulation techniques developed recently, multimode fibers (MMFs) and scattering media have been widely utilized to generate random speckle patterns.  
For example, Wang {\it et al.} proposed fast speckle generation based on a single MMF and a photonic time stretch technique \cite{wang2018ultrafast}.
In that study, a pulse laser was used to produce a short optical pulse with a broadband spectrum, 
and the pulse signal was stretched in the temporal domain 
and converted into speckle patterns via the MMF.
They demonstrated the reconstruction of 
a 27$\times$27-pixel image with 500 speckle patterns.
In addition, Shi {\it et al.} proposed a random speckle encoding scheme based on a hybrid wavelength and spatial phase modulation, which has the potential for high-speed pattern generation of up to gigahertz rates \cite{Shi:21}. 
Furthermore, Fukui {\it et al.} proposed the use of a chip-scale integrated optical phased array for imaging through an MMF \cite{Fukui:21_Lightwave}.  
Using the speckles generated from the method, they obtained fine 2D images of a target object.
These approaches have suggested great potential for accelerating random pattern illumination. However, to the best of our knowledge, the experimental demonstration of high-speed random pattern illumination at GHz rates has not been reported.  
In this study, we propose and experimentally demonstrate an ultrafast method for random 
speckle pattern generation. 
The proposed generation approach is based on MMF speckles induced by 
fast phase modulation at 10 GHz rates, 
which is a five to six orders of magnitude higher than conventional modulation approaches using SLMs.
While a similar modulation approach has been used for compressive sensing of sparse radio-frequency signals \cite{Valley:16}, 
this study represents the first experimental demonstration of high-speed speckle-based information processing. 
In contrast to existing generation schemes that are based on using a pulse-laser as a broadband spectrum source, the proposed approach is simple, versatile, and has the capability of generating a continuous stream of speckle patterns. 
Thus, in principle, information processing without dead time (i.e., the time during which the system cannot operate) is possible.
As a proof-of-concept demonstration of a simple ultrafast random pattern generator, 
we show that 28$\times$28-pixel images of digits (hereafter, digit images) acquired at gigahertz rates 
can be accurately classified using a simple neural network.
This suggests that the proposed generator has the potential to act as a photonic preprocessor for fully photonic information processing.
\section{Multimode fiber speckles}
The proposed random pattern generation is based on dynamic speckles produced from a single and long MMF.
MMFs generally support numerous guided modes with different phase velocities.
When monochromatic coherent light is input to an MMF, 
it can produce a pseudorandom speckle pattern in the intensity distribution at the end facet of the MMF due to the interference between guided modes with different phase velocities. 
Because the phase velocities of the guided modes depend on the frequency (wavelength) of the input light, the speckle pattern is highly sensitive to the input frequency. 
Moreover, the sensitivity of the MMF speckle to the input frequency increases as the MMF length increases \cite{Rawson:80,Redding:13}. 
Accordingly, dynamic speckles can be easily generated with a long MMF by modulating the phase $\phi(t)$ of the input light and inducing an instantaneous input frequency, 
$f(t) \sim d\phi(t)/dt$.
The variety of the generated speckles depends on the instantaneous input frequency.
In our previous work, we have utilized fast optical phase modulation at 12.5 gigasamples per second (GS/s) to induce fast spatiotemporal speckle dynamics in a long MMF for photonic reservoir computing \cite{Sunada:20}.
In this study, we use similar techniques for generating fast random speckle patterns.

\begin{figure}[htbp]
\centering\includegraphics[width=8.5cm]{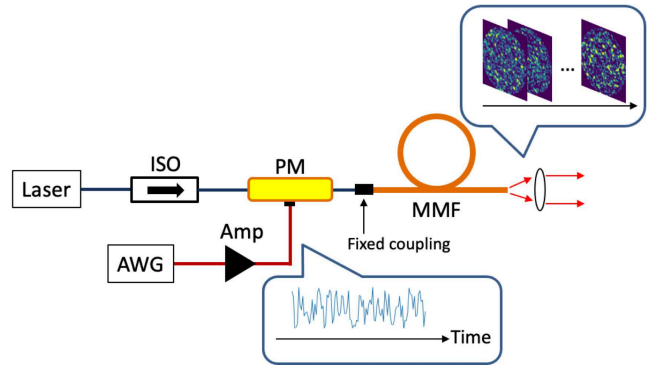}
\caption {\label{fig_specsystem}
Schematic of speckle pattern generation system. ISO represents the optical isolator, Amp represents the electric amplifier, AWG represents the arbitrary waveform generator, PM represents the phase modulator, and MMF represents the multimode fiber. 
A polarization-maintaining single-mode fiber was coupled to the MMF with a standard FC/PC connector.  
}
\end{figure}

Figure~\ref{fig_specsystem} shows the speckle generation system used in our experiment. 
It primarily consists of a coherent light source, a waveform generator, an optical phase modulator, and an MMF. 
In the experiment, a narrow linewidth tunable laser (Alnair Labs, TLG-220, linewidth $<$100 kHz, 30 mW) was used as a coherent light source. 
To dynamically vary the input frequency for the speckle modulation, the laser light was phase-modulated using a lithium niobate phase modulator (EO Space, PM-5S5-20-PFA-PFA-UV-UL, 16 GHz bandwidth) with a uniformly distributed pseudorandom sequence generated using an arbitrary waveform generator (Tektronix, AWG70002A, 25 GS/s).
The pseudorandom sequence can induce a variety of instantaneous input frequencies for generating speckles. 
The modulated light was sent from a polarization-maintaining single-mode fiber to an MMF. 
We used a commercially available step-index multimode fiber with a core diameter of 200 $\mu$m, numerical aperture (NA) of 0.39, and lengths varying from 1 m to 100 m. 
According to a theoretical estimation \cite{goodman2007speckle}, a 100-m MMF is expected to support over $\sim$10$^4$ guided modes at an input wavelength of $\lambda_0 = $1550 nm.

\section{Results}
\subsection{Speckle contrast measurement}
\begin{figure}[htbp]
\centering\includegraphics[width=8.2cm]{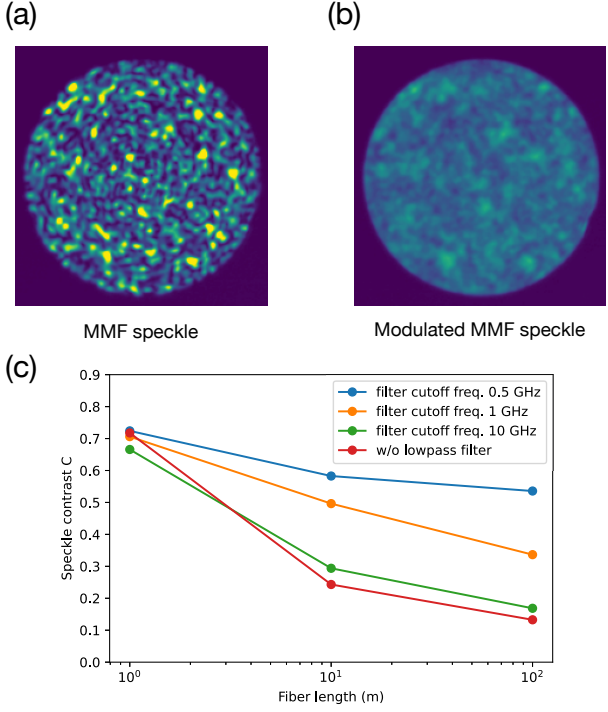}
\caption {\label{fig_lendep}
Measured MMF speckle patterns (a) for non-modulated light input and (b) for phase-modulated light input at 25 GS/s. 
The exposure time of the InGaAs camera was set to 10 ms. 
In (b), the phase-modulated light dynamically induced a number of speckle patterns, and the time-averaged pattern was measured during the exposure time. Thus, the speckle 
contrast $C$ was reduced to 0.13. 
The MMF length was set at 100 m. 
(c) Speckle contrast $C$ as a function of the MMF length. 
To investigate the dependence of the modulation bandwidth on $C$, 
the modulation signal was low-pass filtered using various cut-off frequencies. 
In the experiment, we used a Butterworth filter as the low-pass filter. 
}
\end{figure}

First, we observed the MMF speckle patterns to evaluate the pattern generation capability of the proposed system.
In the experiment, the end facet of the MMF was imaged onto an InGaAs camera (SYNOS, ISA041H2) with a 20$\times$ near-infrared microscope objective (M Plan Apo NIR 20X). 
The exposure time of the camera was set to 10 ms. 
Figure~\ref{fig_lendep}(a) shows a representative example of the measured MMF speckle pattern for an input wavelength of 1550 nm. 
When the input light was phase-modulated at a GHz rate in the system [Fig.~\ref{fig_specsystem}], a number of speckle patterns could be dynamically generated. 
Thus, the time-averaged pattern with a low contrast was measured with the camera [Fig.~\ref{fig_lendep}(b)]. 
Considering that 
the dynamically changing speckle patterns could not be directly captured by a camera with a relatively low frame rate, we used the speckle contrast $C = \sigma/\bar{I}$ (where $\bar{I}$ and $\sigma$ are the mean value and standard deviation of the measured intensity patterns, respectively) to characterize the variety of the generated speckle patterns.    
A value of $C = 1$ indicates that only a single speckle pattern was 
involved with the measured pattern if it was well developed. 
If a number of speckle patterns are involved in the measured pattern, 
$C$ decreases due to the time-averaging during the exposure time of the camera (for instance, $C$ was approximated as $M^{-1/2}$ for $M$ independent speckle patterns with equal intensities \cite{Rawson:80}).
Figure~\ref{fig_lendep}(c) shows $C$ as a function of the MMF length. 
As expected, $C$ decreased as the MMF length increased, suggesting that the sensitivity of the speckle pattern to the length of the MMF could be enhanced \cite{Rawson:80}.
The modulation speed or modulation bandwidth 
is also a crucial parameter for decreasing the speckle contrast
because the input frequency $f(t)$ for dynamic speckle generation
can be characterized as $f(t) = 1/(2\pi)d\phi(t)/dt$.
To investigate the modulation dependence of the speckle contrast, we changed the bandwidth of the pseudorandom signal used for the phase modulation using a low-pass filter and measured the speckle contrast.  
As shown in Fig.~\ref{fig_lendep}(c), as the cutoff frequency of the low-pass filter increased (i.e., as modulation bandwidth increased), $C$ decreased. 
This result suggests that the number of speckle patterns generated in the system is scalable to the MMF length and modulation speed.
In the remaining experiments, we used the 100-m MMF for dynamic speckle generation. 

\subsection{High-speed random speckle projection}
\begin{figure}[htbp]
\centering\includegraphics[width=8.8cm]{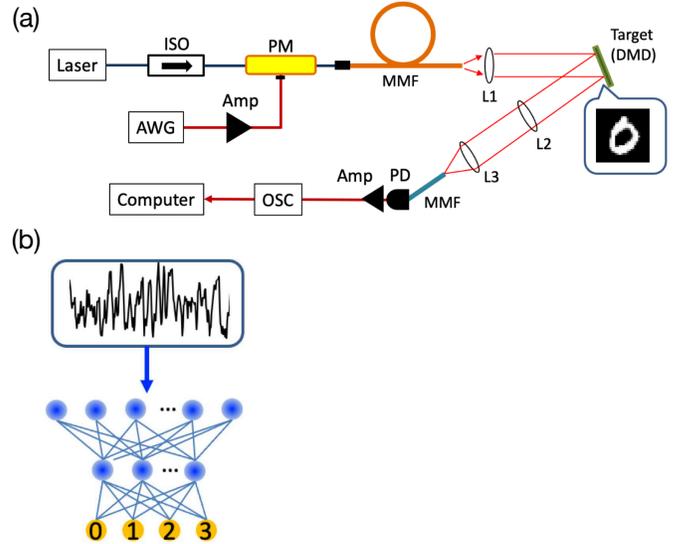}
\caption {\label{fig_expset}
(a) Experimental setup for speckle projection. 
The output from the end facet of the MMF was collimated with a lens (L1).
The target image was displayed on the digital micro mirror device (DMD). 
The reflected light from the DMD was coupled to the receiver fiber (MMF) 
via two lenses (L2 and L3).
PD represents the photodetector, Amp represents the electric amplifier, and OSC represents the digital oscilloscope.
(b) Schematic of neural network for digit recognition. 
The image-encoded time-domain signal was used as an 
input to the network with a fully connected layer 
of 100 units. 
}
\end{figure}

We applied the proposed random speckle projection to encode the image of an object
as a time-domain signal. 
The experimental setup is shown in Fig.~\ref{fig_expset}(a).
The speckle patterns were generated using the MMF 
from the phase-modulated input light at a rate of 12.5 GS/s. The speckle patterns were then 
used as illumination to encode a target image, 
which was displayed on a digital micromirror device (DMD).
The reflected light from the DMD was collected with two lenses, 
coupled to a multimode fiber with a core diameter of 50 $\mu$m, 
and sent to a fast photodetector (Thorlabs, RXM25BF, 500 kHz-25 GHz).
The signal waveform was measured with a digital oscilloscope 
(Tektronix, DPO72504DX, 25 GHz bandwidth, 50 GS/s). 

\begin{figure}[htbp]
\centering\includegraphics[width=8.6cm]{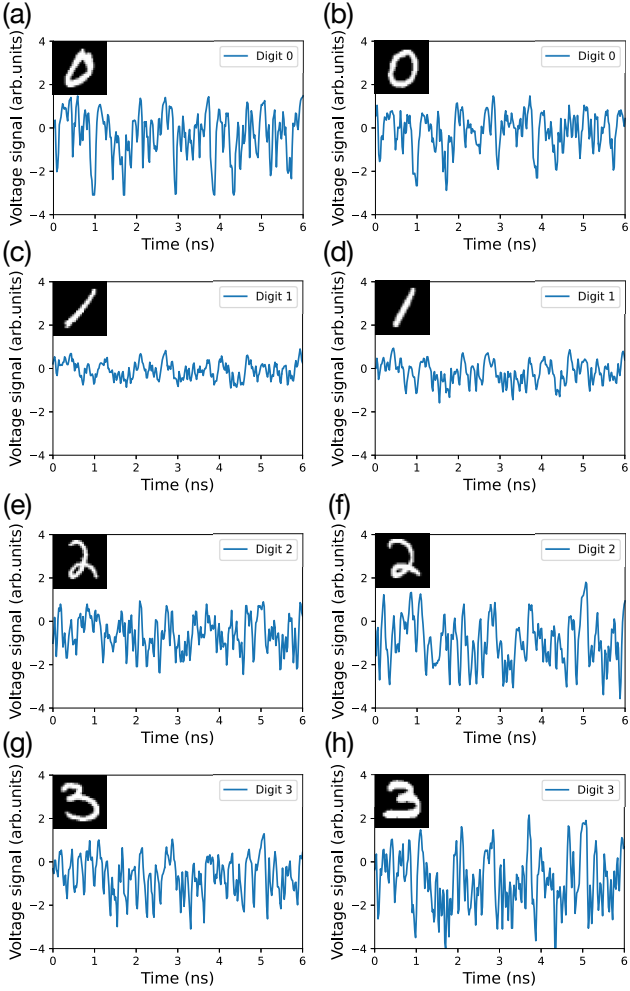}
\caption {\label{fig_time}
Experimental demonstration of the proposed speckle projection. 
(a)--(h) show the time-domain signals (voltage signals acquired by the 
oscilloscope), which encode the image information 
corresponding to each inserted digit image.
The signal waveforms depend strongly on the digit images.   
}
\end{figure}

Figure~\ref{fig_time} shows examples of the 
time-domain signals generated in the experiment, which encoded 
the Modified National Institute of Standards and Technology (MNIST) 
handwritten digit images (``0'',''1'', ``2'', and ``3'') \cite{MNIST_LeCun_url,726791}.
The figure shows that different digit images can be successfully encoded as different 
waveforms, depending on the shapes, locations, and sizes of the digits.
In these examples, the acquisition time $T_N$ was set to 6 ns, 
wherein $N = T_N/\tau_s = $300-pattern-encoded signal was obtained with a sampling time interval of $\tau_s = $ 0.02 ns in the oscilloscope.  

\begin{figure*}[htbp]
\centering\includegraphics[width=11cm]{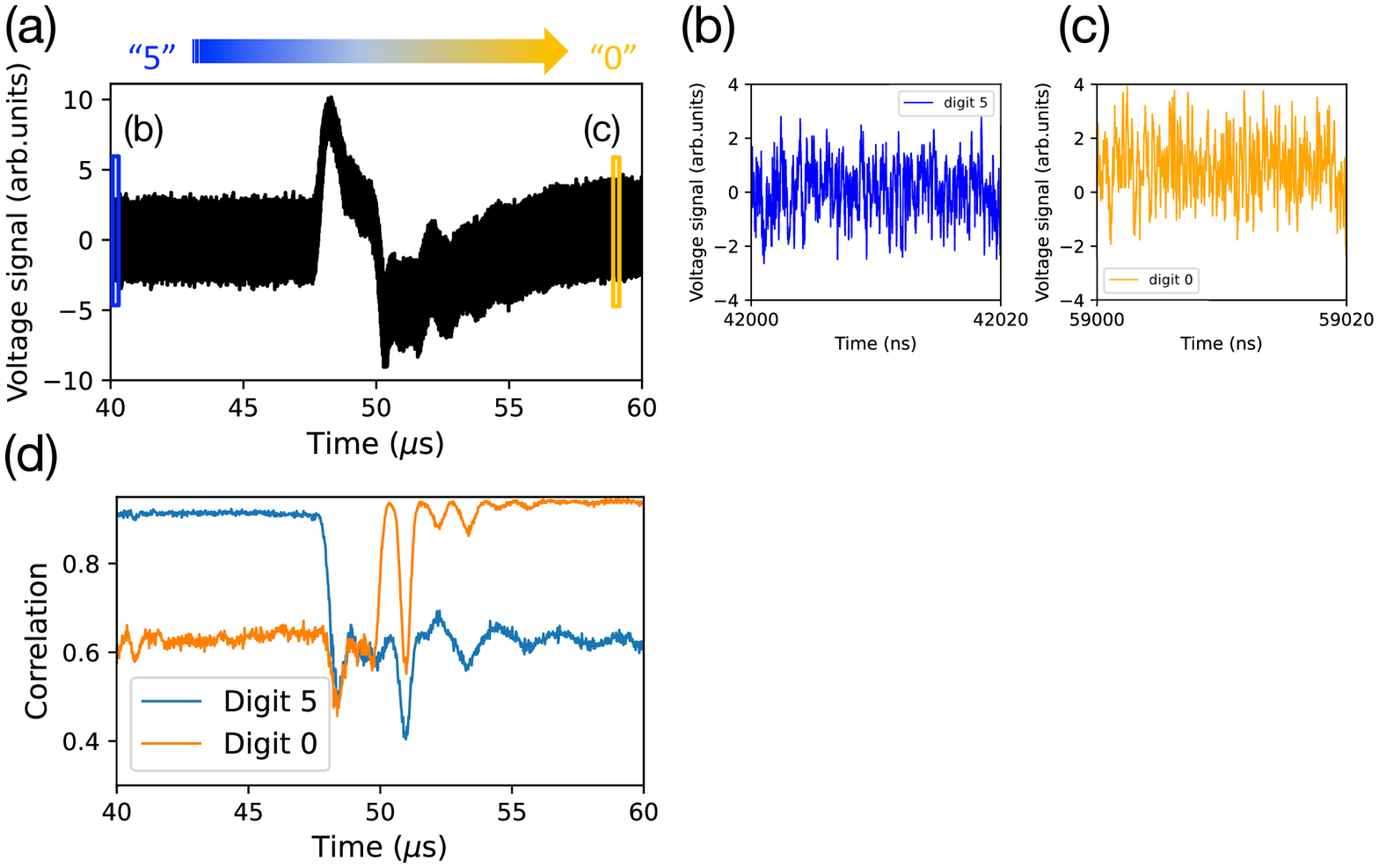}
\caption {\label{fig_switch}
(a) The time-domain signal encoding the switching behavior of the DMD display images from digit ``5'' to digit ``0''.
The switching process occurred around 50 $\mu$s. 
In this demonstration, we set $T_N = 20$ ns for a clear display.  
The waveforms before and after the switching process are shown in (b) and (c), respectively. 
(d) Correlation value $C_i(t_j)$ 
for digit ``$i$'' (i = 0 and 5) as a function of discrete 
time $t_j = jT_N$ ($j = 0,1,\cdots$). 
The correlation analysis clearly reveals the fast switching of the DMD display images from ``5'' to ``0'' with a relaxation oscillation.  
}
\end{figure*}

To demonstrate the capability of capturing dynamic scenes, 
we encoded the transient switching behavior of the images shown on the DMD display 
from digit ``0'' to digit ``5'' as the time-domain signal. 
The result is shown in Fig.~\ref{fig_switch}(a), where
the laser light was repeatedly phase-modulated with the same pseudorandom signal
of $T_N = 20$ ns. 
The switching behavior was analyzed through the correlation 
measurement of the time-domain signal.
We computed the correlation between the signal $I(t)$ shown in Fig.~\ref{fig_switch}(a)
and the signal $I_i(t)$ with encoding digit ''$i$'' ($i =$ 0 or 5) [Figs.~\ref{fig_switch}(b) and \ref{fig_switch}(c)], which is 
defined as $C_{i}(jT_N) = \langle I(t+jT_N)I_i(t)\rangle_{T_N}/(\sigma\sigma_i)$, where $\langle \cdot \rangle_{T_N}$ denotes the time average from $t = 0$ to $t = T_N$, 
and $\sigma$ and $\sigma_i$ denote the standard deviations of $I$ and $I_i$, respectively. 
In Fig.~\ref{fig_switch}(d), the correlation $C_i$ is plotted as a function of discrete time $t_j = jT_N$ $(j=0,1,\cdots)$, where $T_N$ was set to 20 ns. 
Fig. \ref{fig_switch}(d) clearly shows that, when the DMD display was switched, the correlation $C_5$ suddenly decreased and $C_0$ suddenly increased.  
This correlation analysis reveals that the proposed approach 
allows capturing the fast dynamical behavior of the microsecond switching phenomenon. 
Faster dynamical behavior can be captured with our approach by setting a shorter $T_N$. 

\subsection{Application to fast single-pixel image classification}
As mentioned in Section~\ref{sec1}, the proposed random speckle generation can be used for accelerating information processing based on a single-pixel detection. 
While many studies have mostly focused on signal and image reconstruction, 
the reconstruction is frequently not the ultimate goal. 
Because the feature information of an object image is compressively encoded 
into a time-domain signal, the acquired data can be directly utilized 
for classification or recognition.
Such image-free classification can eliminate redundant image acquisition, 
enable higher data efficiency \cite{8918920}, and be used in an application requiring real-time and high-speed control, such as flow cytometry cell sorting \cite{Ota_Science2018}. 
Although many image-free classification schemes based on single-pixel detection have been proposed 
and demonstrated \cite{Zhang:20,Jiao:19,Kumar:21}, this study represents the first demonstration 
of high-speed (potentially at GHz frame rates) image-free classification 
using the proposed speckle generation method. 

The experimental setup was the same as that shown in Fig.~\ref{fig_expset}(a). 
In the experiment, we set the acquisition time $T_N$ to 6 ns to encode a digit image 
with a sampling time interval $\tau_s = $ 0.02 ns ($N = T_N/\tau_s = 300$ points 
was used for the purpose of classification). 
As a classifier, we chose a simple neural network that had 
a single fully connected layer of 100 nodes with a rectified linear unit (ReLU) activation function [Fig.~\ref{fig_expset}(b)].
We did not use convolutional neural network (CNN) layers, which have frequently been used for high-performance recognition, because 
we focused on the effectiveness of the speckle projection approach for classification purposes. 
Additionally, we note that a simple neural network can be implemented in an application-specific circuit for low latency operation.
For the proof-of-concept experiment, we chose 
the MNIST handwritten digit database \cite{MNIST_LeCun_url,726791}, 
and used a subset of the original database consisting of 10000 handwritten 
images of digits from ``0'' to ``3'' as our four-class classification database. 
Each digit image was separated into 9000 (2250$\times$4) images for training 
and 1000 (250$\times$4) images for testing. 
In the training process, cross entropy was chosen as the loss function, 
and it was minimized with the adaptive moment estimation 
gradient descent algorithm (ADAM) \cite{kingma2014adam}. 

\begin{figure}[htbp]
\centering\includegraphics[width=8.8cm]{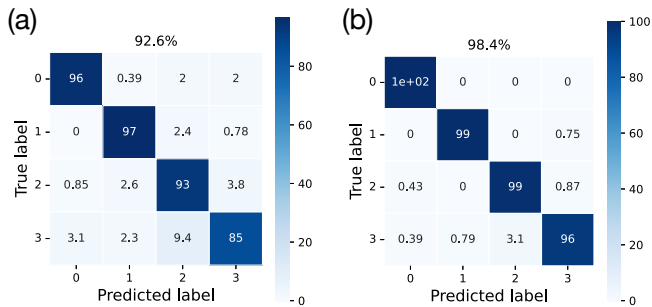}
\caption {\label{fig_confusion}
Performance of the single-pixel image classification for the testing database, as demonstrated by normalized confusion matrices for (a) the experimental results and (b) the numerical results.  
In the experiment, we set the acquisition time $T_N$ to 6 ns (corresponding to $N = 300$, the number of the features used for classification).
In the numerical simulations, we used a Gaussian random matrix as a measurement matrix and generated $N=$300 features for the input to the neural network classifier.
The neural network model for the classification was the same as that used for the experimental results.
The classification accuracies for the experiment and numerical simulations 
were 92.6 $\%$ and 98.4 $\%$, respectively. 
}
\end{figure}

The confusion matrix in Fig.~\ref{fig_confusion}(a) shows 
the distribution of predicted labels for each class. 
Most predicted labels were distributed along the diagonal line and matched the true labels. 
The total classification accuracy was 92.6 $\%$. 
However, a significant error arose as a result of mistaking digit ``3'' for ``2'' and vice versa, 
as shown in the confusion matrix.  
For comparison, we performed numerical simulations, for which 
an $N\times 784$ Gaussian random mask matrix was used to generate $N$ features for a digit image (with 28$\times$28 = 784 pixels), and the same neural network was used for the classifier.  
The result is shown in Fig.~\ref{fig_confusion}(b).
A similar error tendency was observed, but the total accuracy (98.4 $\%$)
was better than that obtained from the experiment. 
This discrepancy may be attributed to the MMF-based speckle generation, 
which was highly sensitive to external environmental fluctuations and measurement errors, particularly quantization errors of 
the 8-bit digital oscilloscope used in the experiment.
See Sec.~\ref{sec_dis} for discussion on the stability of the MMF-based speckle generation. 

\subsection{Wavelength-division multiplexing}
\begin{figure*}[htbp]
\centering\includegraphics[width=11cm]{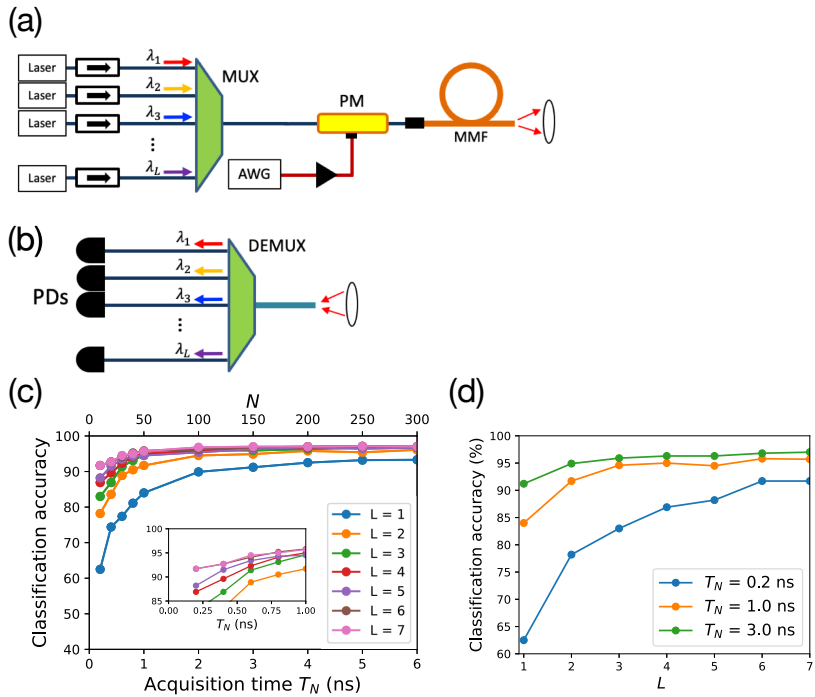}
\caption {\label{fig_wavelen}
Top two panels: schematics of (a) parallel speckle pattern generation system based on wavelength-division multiplexing and (b) measurement system for wavelength-demultiplexing.  
MUX and DEMUX represent the optical multiplexer and optical demultiplexer, respectively. 
Bottom two panels: classification accuracy as a function of (c) acquisition time $T_N$ and 
(d) the number of multiplexed wavelengths $L$. 
For $L \ge 6$ and $T_N >$ 1 ns, the accuracy exceeded 97$\%$, which is 
comparable to the numerical result shown in Fig.~\ref{fig_confusion}(b).
The inset in (c) shows the accuracy in the range of $T_N < $1 ns. 
}
\end{figure*}

As shown in the previous subsection, our approach enables high-speed projection operations at an acquisition rate of $1/T_N \approx 167$ megaframes per second (Mfps). 
For further acceleration of the projection operation, our approach can incorporate the wavelength-division multiplexing scheme, as illustrated in Figs.~\ref{fig_wavelen}(a) and \ref{fig_wavelen}(b). 
To verify the effect of the wavelength-division multiplexing, we used a tunable laser to produce laser light with different wavelengths ranging from 1547 to 1553 nm at 1 nm intervals, and we acquired the image-encoded time-domain signals with these wavelengths instead of the parallel acquisition using a multiplexer and demultiplexer.  
Figures~\ref{fig_wavelen}(c) and \ref{fig_wavelen}(d) show 
the classification accuracy as a function of acquisition time $T_N$ and the number 
of wavelengths $L$, respectively. 
The accuracy improved as $T_N$ and $L$ increased. 
For instance, an accuracy of nearly 95$\%$ was achieved with $L \ge 6$ at the acquisition rate of up to $1/T_N = 1$ Gfps [Fig.~\ref{fig_wavelen}(d)]. 

\section{Summary and discussion \label{sec_dis}}
The main contribution of this work is the experimental demonstration of an ultrafast method for dynamic speckle pattern generation based on high-speed phase modulation. 
The generation of the speckle patterns can be controlled with the phase 
modulation rate.
The speckle generation at over 10 GS/s-modulation rates is much faster than the conventional approach using SLMs.  
We also demonstrated a single-pixel image classification using the proposed speckle pattern generation approach, and we showed that our approach can overcome the major bottleneck of image capture and achieve an acquisition rate of nearly 1 Gfps.
Further improvement is possible by combining this technique with various multiplexing techniques.
We expect that the proposed approach of fast speckle generation is potentially applicable not only for high-speed image classification but also for frame rate improvement of imaging \cite{Higham:2018aa,Xu:18} or dynamic target tracking \cite{Zha:21,SHI2019155,Zha:22}.

Despite the advantages of the proposed approach, there is room for further improvement. 
One such improvement would be better system stability for the speckle pattern generation. 
Because speckle pattern generation utilizes a long MMF, it is highly sensitive to thermal, vibrational, and phonic noises, which can fluctuate the speckle patterns and thereby degrade the classification performance.
One way to minimize these external perturbations is to isolate the MMF from the external environment by placing it inside an isolation box \cite{Vinckier:15}.
Considering that the modal distortions induced in an MMF 
are essentially linear phenomena, they can be compensated for by controlling the 
fiber geometry (i.e., the fiber bending) \cite{Anderson:96}.
Another way is to use a more compact photonic scattering device, such as an on-chip disordered photonic chip \cite{Redding:2013aa}, or the combination of a photonic chip and a shorter MMF instead of solely using a long MMF.

Secondly, it is important to increase the number of independent speckle patterns $M$ generated from the system for applications in image reconstruction.
We roughly estimated $M = C^{-2} \approx 59$ from the speckle contrast shown in Fig.~\ref{fig_lendep}(b). 
Although this estimation is based on the assumption that the speckle patterns were independent and had
equal intensities, the estimated number $M$ is far fewer than expected from the number of the guided modes ($\sim10^{4}$) in the MMF. 
This may be attributed to the fact that (i) 
not all guided modes are excited in the present system 
and/or (ii) the generated speckles are correlated with each other. 
This weakness may be alleviated using an algorithm that imposes no strict limitation on uncorrelated speckles \cite{doi:10.1063/1.5001750}. 
In addition, the wavelength-division multiplexing techniques will be effective 
for the generation of more independent speckle patterns in parallel. 
Therefore, for an input wavelength of $L$, the number of the patterns is $LM$. 
For instance, we can generate $LM = $1200 patterns for $M=60$ and $L = 20$ (with 1 nm intervals ranging from 1540 nm to 1560 nm), which is sufficient for the reconstruction of an MNIST image (with 28$\times$28 pixels).
Further improvement can be made by incorporating a space-division multiplexing using multi-core fiber receivers in the proposed approach. 
The image-encoding based on space- and wavelength-division multiplexing techniques will be reported in a future paper. 

An interesting future study could involve combining the proposed approach with photonic reservoir computing circuits \cite{Vandoorne:2014aa,Sunada:21,Nakajima:2021aa}, 
which enable high-speed processing for time-dependent signals.
In conventional schemes, the image information is electrically converted into time-domain signals with a mask \cite{Nakajima:2021aa}, 
but with our approach, the process can be accomplished optically. 
In other words, the proposed system can be utilized as a photonic preprocessor of reservoir computing for high-speed dynamic scene capture. 
This opens a novel pathway toward the realization of all-optical computing. 

\section*{Acknowledgments}
This work was partly supported by JST PRESTO (JPMJPR19M4),  
JSPS KAKENHI (20H04255),  
Grant for Basic Science Research Projects from the Sumitomo Foundation (210495), and JKA promotion funds from KEIRIN RACE (2022M-208).
%}.

\bibliographystyle{apsrev}
\bibliography{/Users/sunada/My_Refs}

\end{document}